\documentclass[11pt]{article}
\usepackage{graphicx,epsfig}
\usepackage{amsmath,chemarrow}

\begin{document}

\begin{center}
{\LARGE Modeling stochastic Ca$^{2+}$ release from a \\
cluster of IP$_3$-sensitive receptors} \vspace{1cm}

{\large L. Diambra \\ Instituto de F\'{\i}sica de S\~ao Carlos,
Universidade de S\~ao Paulo, \\
Caixa Postal: 369, cep: 13560-970, S\~ao Carlos SP, Brazil. \\
diambra@if.sc.usp.br}
\end{center}

\begin{abstract}
We focused our attention on Ca$^{2+}$ release from the endoplasmic
reticulum through a cluster of inositol 1,4,5-trisphosphate (IP$_3$)
receptor channels. The random opening and closing of
these receptors introduce stochastic effects that have been observed
experimentally. Here, we present a stochastic version of Othmer-Tang
model for IP$_3$ receptor clusters. We address the average behavior
of the channels in response to IP$_3$ stimuli. We found, by
stochastic simulation, that the shape of the receptor response to
IP$_3$ (fraction of open channels versus [IP$_3$]), is affected by
the cytosolic Ca$^{2+}$ level. We also study several aspects of the
stochastic properties of Ca${2+}$ release and we compare with
experimental observations.
\end{abstract}
\parskip=0.5\baselineskip
\newpage

\section{Introduction}

Ionized calcium (Ca$^{2+}$) not only represents the most common
signal transduction element relaying information within cells to
control a wide array of activities including secretion,
contraction and cell proliferation, but also is invariably
involved in cell death (Berridge, 1997). To coordinate all of
these functions, cytosolic Ca$^{2+}$ needs to be precisely
regulated in space, time and amplitude. Normal concentrations of
cytosolic Ca$^{2+}$ ([Ca$^{2+}$] $\sim$ 100 nM) is 20000 fold
lower than the 2 mM concentration found extracellularly. Under
resting conditions this gradient is maintained by active extrusion
of cytosolic Ca$^{2+}$ by Ca$^{2+}$ATPases present in the plasma
membrane and in the endoplasmic reticulum (ER), or in the
sarcoplasmic reticulum in electrically excitable cells. These
pumps counterbalance the leak of Ca$^{2+}$ into the cytosol from
both the extracellular space and the ER Ca$^{2+}$ store.

A wide variety of extracellular stimuli cause the increase of
[Ca$^{2+}$] to exert their effect. In non-excitable cells, this
increase is triggered by inositol(1,4,5)-trisphosphate (IP$_3$),
produced upon activation of phospholipase C (Berridge, 1997).
IP$_3$ rapidly diffuses into the cytosol, where it interacts with
the inositol(1,4,5)-triphosphate receptors (IP$_3$R), to promote
the release of Ca$^{2+}$ into the cytosol. Depending on the cell
type, the resulting cytosolic Ca$^{2+}$ signal can have a complex
spatio-temporal composition (Berridge, 1997). It is generally
recognized that, in the presence of a constant external stimulus,
the Ca$^{2+}$ displays spiking behavior. These Ca$^{2+}$
oscillations can be spatially localized or extended, spreading as
a wave throughout the entire cell.

The process of Ca$^{2+}$ released from the ER through channels is
nonlinear since, increased Ca$^{2+}$ concentration in the cytosol
favors channel opening. This autocatalytic amplification is called
calcium-induced calcium release (CICR). There are a variety of
channels showing CICR. Ca$^{2+}$ release is terminated by closure
of the channels at high Ca$^{2+}$ levels, after which Ca$^{2+}$ is
removed from the cytosol by the action of the Ca$^{2+}$ ATPases.

It has been observed that Ca$^{2+}$ release channels are spatially
organized in clusters. Single release channel, named Ca$^{2+}$
blips, has been observed in experiments (Bootman et al., 1997).
Collective opening and closing of several Ca$^{2+}$ channels in a
cluster, named puffs, have also been observed in experiments
(Callamaras et al., 1998). That suggests a hierarchy of calcium
signaling events from small blips to large puffs (Lipp and Niggli,
1998; Bootman et al., 1997). Improved spatial and temporal
resolution in recording reveal that there is not a clear
distinction between fundamental blips and elementary puffs.
Channels open and close in a stochastic way. Ca$^{2+}$ release by
one channel increase the open probability for the neighboring
channels. Thus at high levels of IP$_3$, neighboring clusters
become functionally coupled by Ca$^{2+}$ diffusion and CICR
support the formation of spatiotemporal patterns of intracellular
Ca$^{2+}$ release.

Watras et al. (1991) found that Ca$^{2+}$ channels in the
cerebellum exhibits four conductance levels that are multiples of
a unit conductance step. This observation suggests that the number
of interacting receptors in one complex can vary from one to four
and supports the hypothesis that the channel is a tetramer.
Examination of the IP$_3$ dependence of channel, yielded Hill
coefficients of 1-1.3 (Watras et al., 1991). However, in
hepatocytes the response to IP$_3$ are positively cooperative with
a Hill coefficient of 3.0-3.4 (Marchant and Taylor, 1997; Dufour
et al., 1997) or. In our numerical simulation we found that the Hill
coefficient increase from 1.15 to 2.2 when total Ca$^{2+}$
(cytosolic + stored) level increase. However, the model
is not able to explain higher values of $n_h$ as that reported by
Marchant and Taylor (1997). This finding suggest that feedback
from cytosolic Ca$^{2+}$ plays a key role in the channel response
to IP$_3$, but is not able to explain high Hill coefficients found in
some tissues.

Experimental finding of Bezprozvanny et al. (1991) clear up
several aspects of the kinetic of the receptors. Thus several
models for Ca$^{2+}$ dynamics in systems involving IP$_3$R have
been proposed (De Young and Keizer, 1992; Othmer and Tang, 1993;
Bezprozvanny and Ehrlich, 1994). These models of IP$_3$R assume a
regulatory site for IP$_3$ on the channel complex, one activating
site and one inhibiting site for Ca$^{2+}$. Experimental findings
suggest that the channel is open when IP$_3$ ion is bound to its
corresponding domain and Ca$^{2+}$ is bound to its activating
domain and not bound to its inhibiting site. In the aforementioned
models, Ca$^{2+}$ binding to the activating regulatory site is a
fast process compared with that of inhibiting binding and all
these models reproduce a bell-shaped curve when one plots the
fraction of open channels as a function of $\log $[Ca$^{2+}$].

De Young and Keizer model (DKM) for IP$_3$R assumes that the
ligands can bind to any unoccupied site on the IP$_3$R
independently of the binding status of the other sites, which
leads to eight possible states of the receptor (De Young and
Keizer, 1992). A more general scheme is to assume state-dependent
binding to each site. In this framework, the simplest model was
developed by Othmer and Tang (1993) which assumes that the binding
order of ions and molecules to the receptor is not free. In the
Othmer-Tang model (OTM) it is assumed that the binding process is
sequential (Othmer and Tang, 1993). Ca$^{2+}$ binds to the
activating site on the receptor only after IP$_3$ has bound, and
that the binding of calcium to inhibitory site occurs only after
calcium is bound to the activating site. This sequential binding
leads to four possible states for the receptor. Bezprozvanny and
Ehrlich (1994) proposed a variation to the OTM which incorporates
an additional closed state corresponding to activating site
occupied which decay rapidly to the open state.

The bindings of IP$_3$ and Ca$^{2+}$ to the regulatory sites are
stochastic events rendering the opening and closing of the channel
a stochastic process. The small number of calcium channels
in a cluster indicates that deterministic models might be insufficient.
In fact, abortive waves cannot be understood in terms of deterministic
models, since in these models an excitation travels steadily if it
travels at all. The stochastic effects are relevant for modeling Ca$^{2+}$
wave propagation. Furthermore, the observation of localized
stochastic Ca$^{2+}$ puffs and the rather small number of channels
creating the localized event, suggest that it is mandatory take into
account the binding processes as stochastic events. The stochastic
dynamics of clustered IP$_3$R has been studied by Swillens et al.
(1999) which model has 14 states. There are also several studies
considering stochastic dynamics focusing on the onset of the
saltatory propagation of calcium waves due to intercluster
diffusion of Ca$^{2+}$ (Keizer et al., 1998; Falcke et al., 2000).
Shuai and Jung (2002b) have been study the statistical properties
of Ca$^{2+}$ release in a stochastic and simplified version of the DKM.

In this paper we regard a Markov-stochastic version of the
OTM to study the stochastic properties of the calcium release of
IP$_3$R clusters. We decided to stick to the Othmer and Tang
approach because the OTM has the least number of channel states
and kinetic parameters, yet it adequately explains the
experimentally observed channel kinetics. After a brief
description of the model in Section 2, we discuss the role of
cytosolic calcium and IP$_3$ in regulating the fraction of open
channels. We show that the fraction of open channels versus IP$_3$
concentration fit a Hill curve which coefficient increase with the
average intracellular calcium concentration.
The amplitude, inter-puff interval (IPI) and size distribution of
calcium puffs are also discussed. Finally, we show that there are
no correlation between two consecutive IPIs, but there are some
correlation between the puff duration and the puff amplitude.

\section{Materials and Methods}
\subsection{Stochastic version of OTM}

In the Othmer and Tang model there is only one subunit/receptor in
a each channel. The three binding sites on the receptor and the
sequential binding scheme allows four different states for the
receptor X$_{i,j,k}$. The index $i$ stands for the IP$_3$ site,
$j$ for the activating Ca$^{2+}$ site, and $k$ for the inhibiting
Ca$^{2+}$. An index is 1 if an ion is bound and 0 if not. The
state transition scheme is given by the pathway
\begin{displaymath}
{\ensuremath{\text{X}_{\text{0,0,0}}}} \ \
\autorightleftharpoons{$k_1$[IP$_3$]}{$k_{-1}$} \ \
{\ensuremath{\text{X}_{\text{1,0,0}}}} \ \
\autorightleftharpoons{$k_2$[Ca$^{2+}$]}{$k_{-2}$} \ \
{\ensuremath{\text{X}_{\text{1,1,0}}}} \ \
\autorightleftharpoons{$k_3$[Ca$^{2+}$]}{$k_{-3}$} \ \
{\ensuremath{\text{X}_{\text{1,1,1}}}},
\end{displaymath}
in which the $k_i$, $i=\pm 1,\pm 2,\pm 3$, are the rate constant
of the state transitions. The channel is only conductive when the
receptor is in the state {\ensuremath{\text{X}_{\text{1,1,0}}}}.
The channels are assumed to be close enough so that Ca$^{2+}$
concentrations can be considered homogeneous throughout the
cluster. We neglect Ca$^{2+}$ diffusion between cluster and
environment without accounting for spatial aspects of formation
and collapse of Ca$^{2+}$ waves. We will not consider calcium
transport across the plasma membrane. Thus the Ca$^{2+}$ simply
flow or is pumped back and forth between the ER and the cytoplasm,
and the total intracellular calcium concentration is constant.
Thus, the dynamic of Ca$^{2+}$ in the cytoplasm is governed by the
equation
\begin{equation}
\frac{d\left[{\rm Ca}^{2+}\right]}{dt}=J_{channel}+J_{leak}-J_{pump},
\end{equation}
where $J_{channel}$ is the calcium flux from the ER to the cytoplasm
through the IP$_3$R channel, $J_{pump}$ is the calcium flux pumped
from the cytoplasm into ER, and $J_{leak}$ is the calcium flux being
leakage from the ER to the cytoplasm. The expression for the
fluxes are given by
\begin{eqnarray}
J_{channel}&=& v_r \gamma _1 \frac{N_{open}}{N} \left( \left[{\rm Ca}^{2+}\right]_{\rm ER} -\left[{\rm Ca}^{2+}\right]\right) \nonumber \\
J_{leak}&=&v_r \gamma _0 \left( \left[{\rm Ca}^{2+}\right]_{\rm ER} -\left[{\rm Ca}^{2+}\right]\right) \nonumber \\
J_{pump}&=&\hat{p}_1 \frac{\left[{\rm
Ca}^{2+}\right]^4}{\left[{\rm Ca}^{2+}\right]^4+\hat{p}_2^4},
\end{eqnarray}
where $v_r$ is the ratio of the ER volume to the cytoplasm volume,
$\gamma _1 $ is the maximal Ca$^{2+}$ fluxes. $N_{open}$ is a
random variable that represent the number of channels open
(receptors in the state {\ensuremath{\text{X}_{\text{1,1,0}}}})
and $N$ is the total number of channels in the receptor. $\gamma
_0$ is the basal permeability of the ER membrane in the absence of
IP$_3$. For the study of Ca$^{2+}$ dynamics in many cell types,
Ca$^{2+}$ exchange with the extracellular medium is much smaller
than the Ca$^{2+}$ flux across the ER membrane (Koster et al.,
1993; Thastrup, 1990). For this reason, Eqs. (1-2) can be
simplified by using the volume average intracellular calcium
concentration [Ca$^{2+}$]$_{ave}=$([Ca$^{2+}$]$+ v_r$
[Ca$^{2+}$]$_{\rm ER}$)$/(1+v_r)$. The [Ca$^{2+}$]$_{ave}$ is a
control parameter whose value we can choose, but that it is not a
dynamical variable. We then define
$C=$[Ca$^{2+}$]/[Ca$^{2+}$]$_{ave}$ and rewrite the Eqs. (1-2) in
the form
\begin{equation}
\dot{C}=\left(1+v_r \right) \left(\gamma_0 +\gamma_1 N
_{open}\right) \left(1-C\right)-p_1 \frac{C^4}{C^4+p_2^4},
\end{equation}
where $p_1 = \hat{p}_1/$[Ca$^{2+}$]$_{ave}$ and
$p_2=\hat{p}_2/$[Ca$^{2+}$]$_{ave}$.

This equation was numerically integrated using fourth order
Runge-Kutta algorithm, with the actual configuration of open
channels $N_{open}$, which is obtained from stochastic simulations
of the kinetics of the receptors. The number of open channels
changes due to six stochastic processes: the IP$_3$ binding
(unbinding) to (from) receptors; the activation (deactivation) of
receptors by binding (unbinding) Ca$^{2+}$ to (from) their
activation domains; the inhibition (activation) of receptors by
binding (unbinding) Ca$^{2+}$ to (from) their inhibition domains.
Transitions involving binding (but not those involving unbinding)
depend on the concentration of IP$_3$ or Ca$^{2+}$. Let us define
the probability distribution vector $\mathcal{P}\left(t
\right)=\left(P_{0,0,0} \left(t \right),
 P_{1,0,0} \left(t \right), P_{1,1,0} \left(t \right),
 P_{1,1,1} \left(t \right)\right)^T$,
where $P_{i,j,k}$ denotes the probability that the receptor is in
the state X$_{i,j,k}$. The dynamic of the vector $
\mathcal{P}\left(t \right)$ is described by the master equation,
\begin{equation}
\dot{\mathcal{P}}= Q\mathcal{P}, \label{me}
\end{equation}
where $Q$ represents a ($4 \times 4$) matrix of transition
probabilities,

\begin{equation}
Q=\left(
\begin{array}{cccc}
1-k_1\left[{\rm IP}_3 \right] & k_{-1} & 0 & 0 \\
k_1\left[{\rm IP}_3 \right]   &  1- k_{-1} - k_{2} [ {\rm Ca}^{2+}]  & k_{-2}  & 0 \\
0 & k_{2} [ {\rm Ca}^{2+}]&   1- k_{-2} - k_{3} [ {\rm Ca}^{2+}]      &   k_{-3}   \\
0 & 0 & k_{3} [ {\rm Ca}^{2+}]  &      1-k_{-3}  \\
\end{array}
\right),
\end{equation}
For the stochastic part of the simulation, the state of each receptors
at time $t+\Delta t$ were independently updated from the previous state
and $Q$. If the time step $\Delta t$ is sufficiently small to allow at
most one transition, the
probabilities for the receptor to transform into the various
possible states within this short time interval are given by
\begin{eqnarray}
P_{{0,0,0} \to {1,0,0}}=k_1\left[{\rm IP}_3 \right]\Delta t \ \ \quad &
P_{{1,0,0} \to {0,0,0}}=k_{-1}\Delta t \nonumber \\
P_{{1,0,0} \to {1,1,0}}=k_2 [ {\rm Ca}^{2+}]\Delta t \quad &
P_{{1,1,0} \to {1,0,0}}=k_{-2}\Delta t \nonumber\\
P_{{1,1,0} \to {1,1,1}}=k_3 [ {\rm Ca}^{2+}]\Delta t \quad &
P_{{1,1,1} \to {1,1,0}}=k_{-3}\Delta t.
\end{eqnarray}
For this model, there is only one possible transition out of the
states X$_{0,0,0}$ and X$_{1,1,1}$, while there are two possible
transitions out of the states X$_{1,0,0}$ and X$_{1,1,0}$. In the
last case a random number $\rho$ between 0 and 1 was drawn from a
uniform distribution for each update step. Then, a channel
initially in state, for example, X$_{1,0,0}$, will be updated to
state X$_{1,1,0}$, if $\rho<P_{{1,0,0} \to {1,1,0}}$, and to state
X$_{0,0,0}$ if $P_{{1,0,0} \to {1,1,0}}< \rho \leq P_{{1,0,0} \to
{0,0,0}}+ P_{{1,0,0} \to {1,1,0}}$, and
remained in its current state otherwise.

All over the paper we have used $\Delta=0.001$ s, and the number
of receptors in a cluster, $N$, was setted to 20. This number is
based on the theoretical predictions of the cluster size by
Swillens et al. (1999), from numerical studies regarding the
requirements of interchannel communication. Furthermore, found
that the optimal cluster size for coherence resonance is around
$N=20$ (Shuai and Jung, 2002a). The values of the parameters for
the kinetic model used in this paper are given in Table 1.

To observe puffs in experiments, calcium diffusion is
suppressed by intracellular loading with the Ca$^{2+}$ buffer EGTA
(Thomas et al., 1998; Marchant et al., 1999; Cheng et al., 1999;
Callamaras and Parker, 2000). With a large loading of EGTA, the
clusters become functionally isolated (Callamaras and Parker,
2000). Under these condition, our model is valid.

\section{Results}
\subsection{The channel response to IP$_3$ and Ca$^{2+}$}

Before to study the statistical properties of the intracellular
Ca$^{2+}$ release, we firstly address the average behavior of
channels in response to IP$_3$ stimuli. In this sense, we simulate
the stochastic dynamics (Eq. 4-5) with the parameters in Table 1
and using [Ca$^{2+}$]$_{ave}$ as control parameter. We compute the
mean value of cytosolic Ca$^{2+}$ concentration in a long time
period (10,000 seconds) as a function of [IP$_3$] for several
values of [Ca$^{2+}$]$_{ave}$ (0.4, 0.8, 1.2, 1.6, 1.9 and 2.4
$\mu$M). The results are sigmoidal curves as is shown in Fig. 1.
It should be noted that the cytosolic mean Ca$^{2+}$ invariably
increases with the level of [Ca$^{2+}$]$_{ave}$. The vertical grey
lines both in Fig. 1 and Fig. 2 are indicating the IP$_3$ levels
for which we have also studied several statistical properties of
the puffs. Further insight can be obtained computing the effect of
the IP$_3$ over the mean fraction of open channels. The mean
fraction of channel that is activated increase monotonically as
the IP$_3$ level increases following Hill curves ($r^2\geq 0.995$
in all cases) as shown in the Fig. 2. The Hill curve is defined by
\begin{equation}
\frac{N_{open}}{N}=V_{max}\frac{[{\rm IP}_3]^{n_h}} {k^{n_h}+[{\rm
IP}_3]^{n_h}}, \label{hill}
\end{equation}
where $k$ is the half maximal value, $V_{max}$ is the maximum
fraction of channels open, and $n_h$ is Hill coefficient. In the
top panel of Fig. 3 we depict the Hill coefficient $n_h$ and in
the bottom panel we depict $k$ and $V_{max}$ for different values
of [Ca$^{2+}$]$_{ave}$. The Hill coefficient increases as
[Ca$^{2+}$]$_{ave}$ increases. At physiological values of
[Ca$^{2+}$]$_{ave}$, the $n_h$ found in our simulation range from
1.15 to 2.2 in very well agreement with the experimental values
which range from 1.0 (Volpe et al., 1990) to greater than 3.7
(Delisle, 1991). In the bottom panel of Fig. 3 we depict the
others two fitted parameters $V_{max}$ and $k$.  The maximum
fraction of open channels $V_{max}$ are between [0.15,0.30] with a
maximum at 0.80 of [Ca$^{2+}$]$_{ave}$, this range is slightly
higher than the values found by Watras et al. (1991), around 0.14
in cerebellar cells (with $n_h=1.3$). It should be noted that the
half-maximal $k$ decrease with the level of [Ca$^{2+}$]$_{ave}$
and range from 0.10 to 0.7 $\mu$M IP$_3$. The half-maximal found
in the Watras experiment was 0.15 $\mu$M IP$_3$.

Bezprozvanny et al. (1991) shown, in experiment where Ca$^{2+}$
channels reconstituted in lipid bilayers, that the fraction of
open cannel as a function of $\log$[Ca$^{2+}$] has an asymmetry
bell-shaped curve, in experiment where calcium was clamped. The
mathematical analysis of the master equation (\ref{me}), allows to
determine the open channel probability (to find the receptor in
the state X$_{1,1,0}$) for a given [Ca$^{2+}$] and [IP$_3$]. The
steady-state distribution vector $\mathcal{P}^{st}$ correspond to
the normalized eigenvector associated to the eigenvalue zero of
the equation of (\ref{me}). Some algebraic manipulations lead to
\begin{equation}
P_{1,1,0}^{st} \left( \left[{\rm IP}_3 \right], [ {\rm Ca}^{2+}]\right)=
\frac{K_1 K_2 \left[{\rm IP}_3 \right] [ {\rm Ca}^{2+}]}
{1+K_1 \left[{\rm IP}_3 \right]+K_1 K_2 \left[{\rm IP}_3 \right]
[ {\rm Ca}^{2+}] +K_1 K_2 K_3\left[{\rm IP}_3 \right] [ {\rm Ca}^{2+}]^2 },
\label{result}
\end{equation}
where $K_1=k_1/k_{-1}$, $K_2=k_2/k_{-2}$ and $K_3=k_3/k_{-3}$.
Fig. 4 depicts the open channel probability for three different
values of IP$_3$ concentration as a function of [Ca$^{2+}$]. The
result (\ref{result}) predicts the bell-shaped curve obtained
experimentally by Bezprozvanny et al. (1991), with the symmetry
axe correctly shifted to left. It should be noted the difference
between Eq. (\ref{hill}) and (\ref{result}). The first one can be
understood as a open channel probability in physiological
conditions described by Eq. (1-2). Thus the fraction of open
channels is computed as the average of fraction of open channels
over a period of time (10,000 seconds in this paper), where
calcium is released (removed) to (from) cytosol many times. On the
other hand, Eq. (\ref{result}) is a steady state probability for a
fix level of calcium and IP$_3$. This situation emulate the
Bezprozvanny et al. experimental conditions, where the calcium was
clamped.

\subsection{Statistical properties of puffs}

We now study the statistical properties of puffs obtained with the
stochastic version of the OTM, the deterministic version of OTM is
not able to address these properties. It is known that IP$_3$
stimuli evokes repetitive spikes in the intracellular [Ca$^{2+}$]
rather than simply increase [Ca$^{2+}$]. In the Fig. 5, we can see
the typical spiking behavior of calcium release by a cluster with
20 channels ($N=20$) and some statistical properties of the
observed puffs 10,000 seconds of simulations. In this case, we
performed the simulation setting [Ca$^{2+}$]$_{ave}$=1.6 $\mu$M
and IP$_3$=0.35 $\mu$M. At the top panel of Fig. 5 we depict the
first 100 seconds of the temporal course of the intracellular
calcium concentration. At the bottom-left panel, the scatter plot
of IPI$_i$ versus IPI$_{i+1}$ shown that there is not temporal
correlation between two consecutive IPIs. At the bottom-center
panel: scatter plot of puff amplitude versus puff duration shown
that there is correlation between puff amplitudes and durations
($r^2$=0.84). Small puff amplitudes are correlated with brief
puffs, the correlation decreases as amplitude puff increases.
Experimental observations of Thomas et al. (1998) indicate similar
correlations, but with a smaller linear regression coefficient
($r^2$=0.69). At the bottom-right panel we shown the puff size
distribution, it should be noted that there are Ca$^{2+}$ releases
of all sizes. All plots in the bottom panel of Fig. 5, and
hereafter, were computed with puffs which amplitudes exceed the
threshold 0.20 $\mu$M, in order to avoid the background
fluctuation. In this paper, the puff size is defined as the area
between the [Ca$^{2+}$] time course and the threshold.

It was experimentally established that when the IP$_3$
concentration increases, the mean frequency of spiking raises, but
the amplitude remain essentially constant. Fig. 6 displays the
mean frequency of puffs as a function of the IP$_3$ concentration
for the some values of [Ca$^{2+}$]$_{ave}$ showed in Figs. 1-2
(0.8, 1.2, 1.6, 1.9 and 2.4 $\mu$M) obtained by stochastic
simulations. We can observe that the onset of spiking behavior in
response to the IP$_3$ stimuli depends on the [Ca$^{2+}$]$_{ave}$.
For high levels of calcium, even low levels of IP$_3$ are able to
produce repetitive puffs. These responses are well characterized
by Hill curves. The Hill coefficients associated to these curves
were essentially independent of [Ca$^{2+}$]$_{ave}$ level. For
[Ca$^{2+}$]$_{ave}$=2.4 $\mu$M the Hill coefficient was 2.1 and
decrease to 1.8 for [Ca$^{2+}$]$_{ave}$=0.8 $\mu$M.

Furthermore, the number of channel recruited in the puffs varies
stochastically from puff to puff. Important, and experimentally
recorded, characteristics of these variabilities are amplitude,
inter-puff interval, duration and size distributions (Bootman et
al., 1997; Sun et al., 1998; Thomas et al., 1998; Callamaras and
Parker 2000; Marchant and Parker 2001). The shape of the puff
amplitude distribution depends on the concentration of IP$_3$ and
the [Ca$^{2+}$]$_{ave}$. In Fig. 7 we shown a diagram of the puff
amplitude distributions in the [Ca$^{2+}$]$_{ave}$-IP$_3$ plane.
In order to comparison the scale of the axis are the same for all
plots in the diagram, between [0.20,1.80] $\mu$M for the
amplitudes of puffs (horizontal axe), and between [0,6000] for the
number of events (vertical axe). No events (NE) were recorded for
[IP$_3$]= 0.07 $\mu$M and low level of [Ca$^{2+}$]$_{ave}$=0.8
$\mu$M). For lower [Ca$^{2+}$]$_{ave}$, the amplitudes of the
spontaneous puffs are typically smaller than 0.20 $\mu$M and are
regarded as fluctuations. For [Ca$^{2+}$]$_{ave}$=0.8 $\mu$M, or
[IP$_3$]= 0.07 $\mu$M, monotonically decreasing amplitude
distribution are found. In the other cases a two-peak amplitude
distributions are mainly found. The characteristic amplitudes
depend on [Ca$^{2+}$]$_{ave}$, but not on [IP$_3$]. For
[Ca$^{2+}$]$_{ave}$=1.6 $\mu$M, the typical amplitude was around
0.80 $\mu$M. The shape of the puff amplitude distributions shown
in the Fig. 7 differ from those observed in experiments, which
have one-peak shape (Sun et al., 1998; Thomas et al., 1998;
Marchant and Parker,2001).

Another important characteristic of the calcium dynamics is the
distribution of time interval between two consecutive puffs.
Recent experimental investigation with {\it Xenopus} oocytes has
revealed that IPI distribution exhibits a single peak. In Fig. 8
we display another diagram in the [Ca$^{2+}$]$_{ave}$-IP$_3$
plane, but now with IPI distributions. In this diagram the range
of the vertical axis are the same for all distributions, between
[0,6000] events. While the ranges of IPI (horizontal axis) are
between [0,80] seconds for the first column ([Ca$^{2+}$]$_{ave}$
=0.8 $\mu$M), and between [0,20] seconds for all other cases. For
[Ca$^{2+}$]$_{ave}$ =0.8 $\mu$M, the IPI distribution seems to
decay exponentially. However, when the level of
[Ca$^{2+}$]$_{ave}$ is increased, skewed single-peak distribution
characterize the IPI distributions. The position of the single
peak range from 1 to 2 seconds increasing with the
[Ca$^{2+}$]$_{ave}$. The shape of the distributions shown in the
Fig. 8 are consistent with observed by Marchant et al. (1999), but
with a typical IPI smaller than that observed in the experiment
(around 2.5 seconds). It is apparent that for high values of
[IP$_3$] and [Ca$^{2+}$]$_{ave}$, the IPI are quite regular.

We also want to illustrate the typical temporal behavior and some
statistical properties, as we shown in the Fig. 5, for two
different situations: i) high [IP$_3$], low [Ca$^{2+}$]$_{ave}$
(Fig. 9); and ii) low [IP$_3$], high [Ca$^{2+}$]$_{ave}$ (Fig.
10). It should be noted that different scale axis were used. In
Fig. 9, where [IP$_3$]=21 $\mu$M, and [Ca$^{2+}$]$_{ave}$=0.8
$\mu$M, we found the distribution of puffs size decay
exponentially, and there are very few great calcium release events
(bottom-right panel). More broad distributions of IPI are yielded
for this case than those shown in bottom-left panel of Figs. 5 and
10. There are also less IPI around 1-2 seconds (bottom-left panel)
than those shown in Fig. 5 or in Fig. 10. The correlation between
puff amplitude and duration seems to be similar than that shown in
bottom-right panel of Fig. 5. When increase the
[Ca$^{2+}$]$_{ave}$ until 2.4 $\mu$M and decrease the [IP$_3$] to
0.07 $\mu$M (Fig. 10), we found a slower exponential decay in the
distribution of puffs size than in Fig. 9 (bottom-right panel). In
this case the distributions of IPI (bottom-center panel of Fig.
10) looks like those shown in bottom-left panel of Figs. 5. Puff
with small amplitude seems to be correlated with short duration.

\section{Discussion and conclusion}

We have developed a stochastic version of the Othmer-Tang model to
study dynamical and statistical properties of Ca$^{2+}$ release of
clusters of IP$_3$-sensitive receptors. In comparison to others
stochastic models (Falcke et al., 2000; B\"ar et al., 2000;
Swillens et al., 1999), our receptor model is simpler and is
represented by only four states. In this stochastic clustered
IP$_3$R model, the Ca$^{2+}$ diffusion between the cluster and the
environment is ignored so that an isolated cluster can be
discussed. However, the channels are assumed to be close enough
and the instantaneous Ca$^{2+}$ diffusion within a cluster is so
fast that the calcium concentration within a cluster is always
homogeneous.

We found that the model reproduce experimental result by Watras et
al. (1991) where it is shown that the channel response to the
IP$_3$ follow a Hill curve. In intact cells, the long latency and
subsequent sigmoidal increase in cytosolic [Ca$^{2+}$] after
addition of agonists (Berridge, 1997), are consistent with either
positively cooperative activation of IP$_3$Rs by IP$_3$ alone or
by feedback from cytosolic [Ca$^{2+}$]. Distinguishing between
these mechanisms, however, is difficult in intact cells, in which
IP$_3$ receptor activation inevitably increases cytosolic
[Ca$^{2+}$]. In broken cells, the use of fluorescent indicators to
measure rapid rates of Ca$^{2+}$ efflux is prone to similar
difficulties. Results from a variety of tissues using superfusion
(Finch et al., 1991), bilayer recording (Watras et al., 1991;
Besprozvanny et al., 1991), fluorescent indicators (Champeil et
al., 1989) or flash photolysis of caged IP$_3$ in intact cells
(Parker et al., 1996) suggest that responses to IP$_3$ are either
not positively cooperative (Finch et al., 1991; Parker et al.,
1996; Watras et al., 1991) or only slightly so (Besprozvanny et
al., 1991; Champeil et al., 1989). Each approach has its
limitations and together the results provide no clear indication
of whether responses to IP$_3$ are positively cooperative in the
absence of positive feedback from cytosolic Ca$^{2+}$. In our
numerical simulation we found that, even in absence of
cooperativity in the model, the Hill coefficient increase from
1.15 to 2.2 when [Ca$^{2+}$]$_{ave}$ increase. However, the model
is not able to explain higher values of $n_h$ as that reported by
Marchant and Taylor (1997). This results suggest that feedback
from cytosolic Ca$^{2+}$ plays a key role in the channel response
to IP$_3$, but to explain high values of $n_h$ found in some
tissues, a cooperative mechanism could be mandatory.

Other important aspect is the frequency encoding. It is well known
that IP$_3$ stimuli evoke repetitive spikes in the intracellular
Ca$^{2+}$ concentration rather than simply elevating the level of
Ca$^{2+}$. Moreover, when the [IP$_3$] is increased, the mean
frequency of spiking raises, but the amplitude remains essentially
constant. Thus a continuous level of stimuli signal is converted
into a frequency-encoded signal: the number of spikes per time
unit (Tang and Othmer, 1995). A common problem found in this kind
of converter devices is the competition between two desirable
goals: i) high sensitivity, the system ideally should be able to
detect even weak signals; and ii) large dynamic range, the system
should not saturate over various orders of magnitude of input
intensity. For this reason the Hill coefficient (when the response
is well-fitted by a Hill curve) its desirable to be less than 1.
In the results derived from our numerical simulation, we can
observe that the mean frequency of the spiking behavior in
response to the IP$_3$ stimuli is well characterized by Hill
curves. The Hill coefficient, between 1.8-2.1, was essentially
independent of [Ca$^{2+}$]$_{ave}$. This value is too high given
narrow dynamical range in order to frequency-encode the IP$_3$
signal. However, this narrow dynamical range observed in isolated
cluster, could be translates into a wider dynamical range as
result of the collective phenomenon by considering an intercluster
communication mechanism. This interesting discussion is the
subject of our current research and will be discussed in a
forthcoming paper.

The small number of the IP$_3$Rs in a cluster introduce stochastic
oscillation into the calcium release dynamics. Experimental data
suggest that the localized calcium release varies in a continuous
fashion due to stochastic variation in both numbers of channels
recruited and durations of channel openings. It is not obvious
which aspects of Ca$^{2+}$ puffs are originally determined by the
dynamics of the Ca$^{2+}$ channels, which properties are
determined by the diffusion and Ca$^{2+}$ binding kinetics, and
which properties are induced from the measurement procedures
(Cheng et al., 1999; Izu et al., 1998). Although in several
experiments one-peak amplitude distributions were observed, Cheng
et al., (1999) suggested that the original calcium puffs should
have an exponential decaying decreasing amplitude distribution.
Our stochastic simulations shown that the amplitude distributions
varies in a continuous fashion from exponential decay to two-peak
distributions. Even for a small threshold (as 0.20 $\mu$M) a peak
is observed in the amplitude distributions for high values both of
[IP$_3$] and [Ca$^{2+}$]$_{ave}$. However, the puffs size (i.e.
amplitude $\times$ duration) distributions have essentially an
exponential decay distribution over the range studied here. These
results indicate that a fixed puff morphology which is sometimes
assumed in literature (Pratusevich and Balke, 1996; Izu et al.,
1998; Cheng et al., 1999) is not a good assumption for Ca$^{2+}$
puff analysis. Furthermore, in our markovian simulation, we found
that the amplitude of puff is correlated with its duration in
agreement with experimental observation reported by Thomas et al.,
(1998).

Other useful way to characterize the calcium puffs dynamics is
using the IPI distributions. The stochastic simulation of the OTM
shown that the IPI distributions varies from exponential decay to
a skewed bell-shape distributions. For physiological values of
IP$_3$ and [Ca$^{2+}$]$_{ave}$ the distributions have a maximum
around 1.5 seconds. The shape is consistent with distribution
observed by Marchant et al. (1999), however, the typical IPI
observed in the experiment, around 2.5 seconds, was slightly
higher. In contrast to the deterministic models, we shown that two
consecutive IPI are not correlated. For this reason, the random
spiking behavior or stochastic oscillations have an essentially
different nature that the periodic oscillations generated by
deterministic models.

We want to remark that the stochastic oscillations are different
from the stochastic excitability discussed by Keizer and Smith
(1998). For the stochastic excitability, once [Ca$^{2+}$] randomly
becomes larger than a threshold, a fast release
(action-potential-like) of [Ca$^{2+}$] followed by a refractory
period can be observed. For the stochastic oscillation studied
here there is not such a threshold. More broad distributions of
puff amplitudes and IPI are yielded for stochastic oscillation.

\section*{Acknowledges}
This research was partially supported by the Deutsche
Forschungsgemeinschaft. The author thanks to Gerhard Thiel and
Nara Guisoni for valuable suggestions. L.D. is grateful for the
hospitality of Professor Friedemann Kaiser and Nonlinear Dynamics
Group at the Darmstadt University of Technology where these ideas
were developed.

\section*{References}

\noindent B\"ar, M., M. Falcke, H. Levine, and L. S. Tsimring. 2000.
Discrete stochastic modeling of calcium channel dynamics. {\it
Phys. Rev. Lett.} 84:5664-5667.

\noindent Berridge, M. J., 1997. Elementary and global aspect of
calcium signaling. {\it J. Physiol.} 499:291-306.

\noindent Bezprozvanny, I., J. Watras, and B. Ehrlich. 1991.
Bell-shaped calcium response curves of Ins(1,4,5)P$_3$- and
calcium-gated channels from endoplasmic reticulum of cerebellum.
{\it Nature.} 351:751-754.

\noindent Bezprozvanny, I., and B. E. Ehrlich. 1994. Inositol
1,4,5-trisphopate (IP$_3$)-gated Ca channels from cerebellum:
conduction properties for divalent cations and regulation by
intraluminal calcium. {\it J. Gen. Physiol.} 104:821-856.

\noindent Bootman, M., E. Niggli, M. Berridge, and P. Lipp. 1997.
Imaging the hierarchical Ca$^{2+}$ signaling system in HeLa cells.
{\it J. Physiol.} 499:307-314.

\noindent Callamaras, N., J. S. Marchant, X. Sun, and I. Parker.
1998. Activation and co-ordination of InsP$_3$-mediated elementary
Ca$^{2+}$ events during global Ca$^{2+}$ signals in {\it Xenopus}
oocytes. {\it J. Physiol.} 509:81-91.

\noindent Callamaras, N., and I. Parker. 2000. Phasic
characteristic of elementary Ca$^{2+}$ release sites underlies
quantal responses to IP$_3$. {\it EMBO J.} 19:3608-3617.

\noindent Champeil, P., L. Combettes, B. Berthon, E. Doucet, S.
Orlowski, M. Claret. 1989. Fast kinetics of calcium release
induced by myo-inositol trisphosphate in permeabilized rat
hepatocytes. {\it J. Biol. Chem.} 264:17665-17673.

\noindent Cheng, H., L. Song, N. Shirokova, A. Gonzalez, E. G.
Lakatta, E. Rios, and M. D. Stern. 1999. Amplitude distribution of
calcium sparks in confocal images: theory and studies with an
automatic detection method. {\it Biophys. J.} 76:606-617.

\noindent Delisle, S. 1991. The four dimensions of calcium
signaling in {\it Xenopus} oocytes. {\it Cell Calcium.}
12:217-227.

\noindent De Young, G. W., and J. Keizer. 1992. A single-pool
inositol 1,4,5- trisphosphate-receptor-based model for
agonist-stimulated oscillations in Ca$^{2+}$ concentration. {\it
Proc. Natl. Acad. Sci. USA.} 89:9895-9899.

\noindent Dufour, J. F., I. M. Arias, and T. J. Turner. 1997.
Inositol 1,4,5-trisphopate and calcium regulate the calcium
channel function of the hepatic inositol 1,4,5-trisphopate
receptor. {\it J. Biol. Chem.} 272:2675-2681.

\noindent Falcke, M., L. Tsimring, and H. Levine. 2000. Stochastic
spreading of intracellular Ca$^{2+}$ release. {\it Phys. Rev. E.}
62:2636-2643.

\noindent Finch, E.A., T.J. Turner, S. M. Goldin. 1991. Calcium as
a coagonist of inositol 1,4,5-trisphosphate-induced calcium
release. {\it Science.} 252:443-446.

\noindent Izu, L. T., W. G. Wier, and C. W. Balke. 1998.
Theoretical analysis of the Ca$^{2+}$ spark amplitude
distribution. {\it Biophys. J.} 75:1144-1162.

\noindent Keizer, J., and G. D. Smith. 1998. Spark-to-wave
transition: saltatory transmission of calcium waves in cardiac
myocytes. {\it Biophys. Chem.} 72:87-100.

\noindent Keizer, J., G. D. Smith, S. Ponce-Dawson, and J. E.
Pearson. 1998. Saltatory propagation of Ca$^{2+}$ waves by
Ca$^{2+}$ sparks. {\it Biophys. J.} 75:595-600.

\noindent Koster, H., C. van Os, and R. Bindels. 1993. Ca$^{2+}$
oscillations in the rabbit renal cortical collecting system
induced by Na$^+$ free solutions. {\it Kidney Int.} 43:828-836.


\noindent Lipp, P., and E. Niggli. 1998. Fundamental calcium
release events revealed by two-photon excitation photolysis of
caged calcium in guinea-pig cardiac myocytes. {\it J. Physiol.}
508:801-809.

\noindent Marchant, J. S., and C. W. Taylor. 1997. Cooperative
activation of IP$_3$ and Ca$^{2+}$ safeguards against spontaneous
activity. {\it Current Biology} 7:510-518.

\noindent Marchant, J. S., N. Callamaras, and I. Parker. 1999.
Initiation of IP$_3$-mediated Ca$^{2+}$ waves in {\it Xenopus}
oocytes. {\it EMBO J.} 18:5285-5299.

\noindent Marchant, J. S., and I. Parker. 2001. Role of elementary
Ca$^{2+}$ puffs in generating repetitive Ca$^{2+}$ oscillations.
{\it EMBO J.} 20:65-76.


\noindent Othmer, H. G., and H. Tang. 1993. In Oscillations and
waves in a model of calcium dynamics. {\it In} Experimental and
Theoretical Advances in Biological Pattern Formation. H. Othmer,
J. Murray, and P. Maini, editors. Plenunm Press, London. 295-319,

\noindent Parker, I., Y. Yao, V. Ilyin. 1996. Fast kinetics of
calcium liberation induced in {\it Xenopus} oocytes by
photoreleased inositol trisphosphate. {\it Biophys J.} 70:222-237.

\noindent Pratusevich, V. R., and C. W. Balke. 1996. Factors
shaping the confocal image of the calcium spark in cardiac muscle
cells. {\it Biophys. J.} 71:2942-2957.


\noindent Shuai, J. W., and P. Jung. 2002a. Optimal intracellular
calcium signaling. {\it Phys. Rev. Lett.} 88:68102-1-68102-4.

\noindent Shuai, J. W., and P. Jung. 2002b. Stochastic properties
of Ca$^{2+}$ release of inositol 1,4,5-trisphosphate receptor
cluster. {\it Biophys. J.} 83:87-97.

\noindent Sun, X., N. Callamaras, J. S. Marchant, and I. Parker.
1998. A continuum of InsP$_3$-mediated elementary Ca$^{2+}$
signaling events in {\it Xenopus} oocytes. {\it J. Physiol.}
509:67-80.

\noindent Swillens, S., G. Dupont, L. Combettes, and P. Champeil.
1999. From calcium blips to calcium puffs: theoretical analysis of
the requirements for interchannel communication. {\it Proc. Natl.
Acad. Sci. USA.} 96:13750-13755.

\noindent Tang, Y., and H. Othmer. 1995. Frequency encoding in
excitable systems and its application to calcium oscillations.
{\it Proc. Natl. Acad. Sci. USA.} 92:7869-7873.

\noindent Tang, Y., J.L. Stephenson, and H. Othmer. 1996.
Simplification and analysis of the models calcium dynamics based
on IP$_3$-sensitive calcium channel kinetics. {\it Biophis. J.}
70:246-263.

\noindent Thastrup, O., P. Cullen, B. Drobak, M. Hanley, and A.
Dawson. 1990. Thapsigargin,  A tumor promoter, discharges
intracellular Ca$^{2+}$ stores by specific inhibition of the
endoplasmic reticulum Ca$^{2+}$ ATPase. {\it Proc. Natl. Acad.
Sci. USA.} 87:2466-2470.

\noindent Thomas, D., P. Lipp, M. J. Berridge, and M. D. Bootman.
1998. Hormone-evoked elementary Ca$^{2+}$ signals are not
stereotypic, but reflect activation of different size channel
clusters and variable recruitment of channels within a cluster.
{\it J. Biol. Chem.} 273:27130-27136.

\noindent Volpe, P., B. H. Alderson-Lang, and G. A. Nickols. 1990.
Regulation of inositol 1,4,5-trisphospate-induced Ca$^{2+}$
release. I. Effect of Mg$^{2+}$. {\it American J. of Physiology.}
258:C1077-C1085.

\noindent Watras, J., I. Bezprozvanny, and B. E. Ehrlich. 1991.
Inositol 1,4,5-trisphopate-gated channels in cerebellum: presence
of multiple conductance states. {\it J. Neurosci.} 11:3239-3245.


\newpage

\section*{Tables and Figures}

\begin{table} [h]
\begin{tabular}{|ll||ll|} \hline
\multicolumn{2}{|c||}{\underline{Kinetic parameters}} &
\multicolumn{2}{|c|}{\underline{Other parameters}}\\
name&value& name& value \\ \hline
$k_{1}$  & 12.0 $(\mu$M$\times$ s$)^{-1}$ & $v_r$         & 1.85  \\
$k_{-1}$ & 8.0  s$^{-1}$ & $\gamma _{0}$  & 0.1  s$^{-1}$\\
$k_{2}$  & 23.4 $(\mu$M$\times \ s)^{-1}$ & $\gamma _{1}$ & 20.5 s$^{-1}$\\
$k_{-2}$ & 1.65 s$^{-1}$ & $\hat{p}_1$    & 8.5 $\mu$M$\times $ s$^{-1}$ \\
$k_{3} $ & 2.81  $(\mu$M$\times \ s)^{-1}$ & $\hat{p}_2$   & 0.065 $\mu$M\\
$k_{-3}$ & 0.21 s$^{-1}$ & $N$            & 20 \\
\hline
\end{tabular}
\caption{Parameters values for the OTM.}
\end{table}


\vspace{3cm}
\begin{figure}[ht]
\includegraphics[width=15cm]{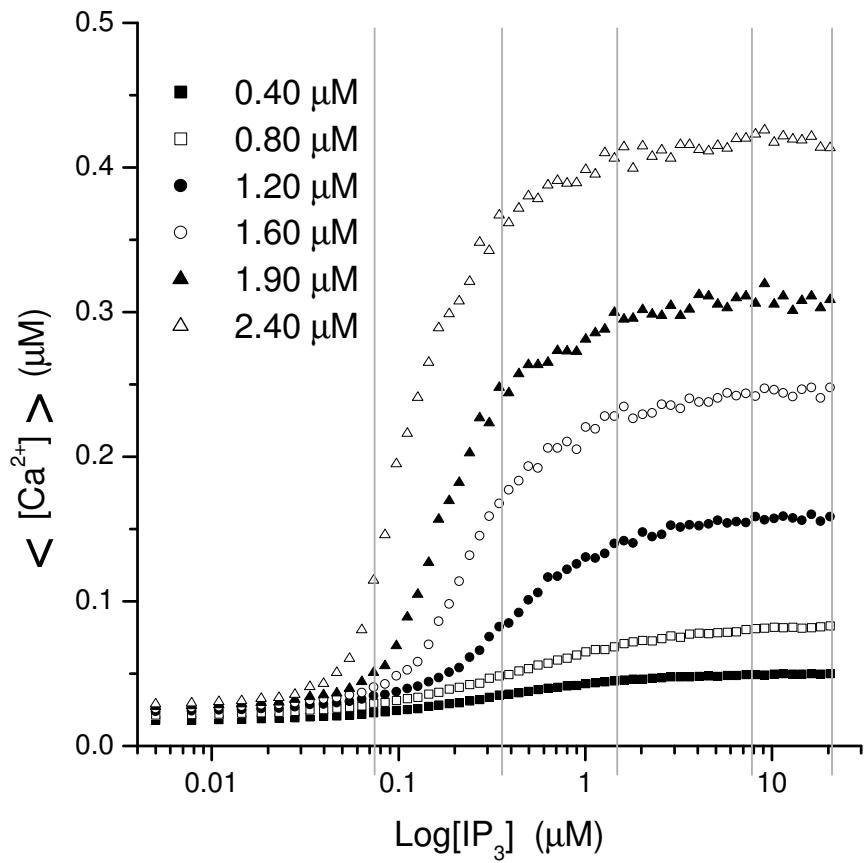} \caption{Mean value
of [Ca$^{2+}$] versus IP$_3$ concentration, for six different
concentrations of the average intracellular calcium
[Ca$^{2+}$]$_{ave}$, computed over 10,000 seconds.}
\end{figure}

\begin{figure}[ht]
\includegraphics[width=15cm]{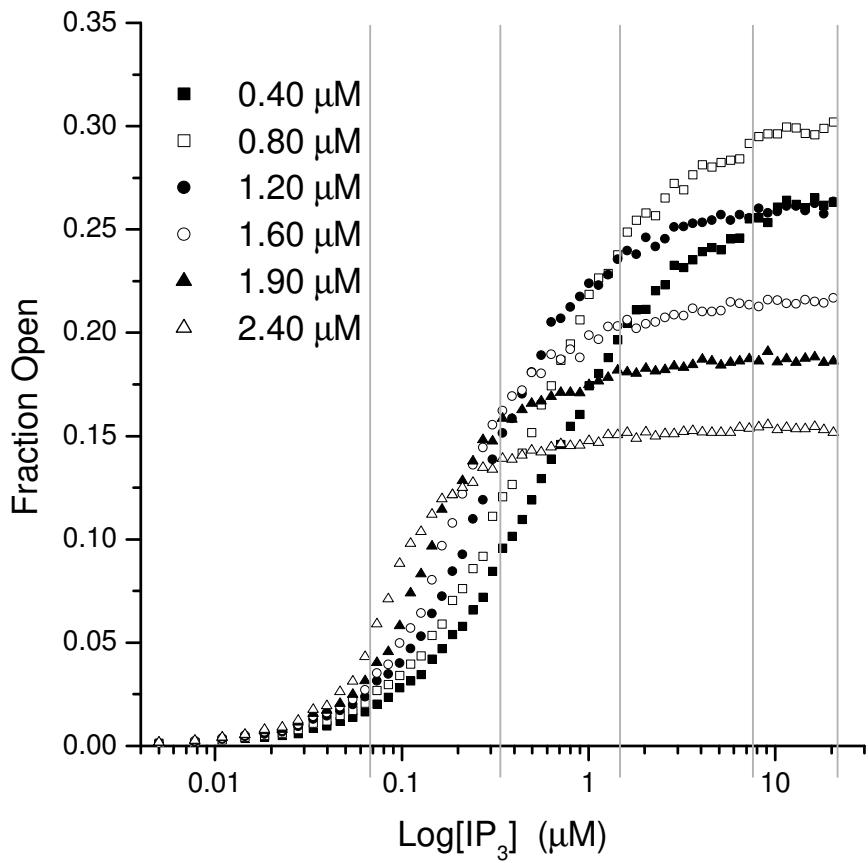}
\caption{Fraction of open channels versus IP$_3$ concentration,
for six different concentrations of the average intracellular
calcium [Ca$^{2+}$]$_{ave}$, computed over 10,000 seconds.}
\end{figure}

\begin{figure}[ht]
\includegraphics[width=10cm]{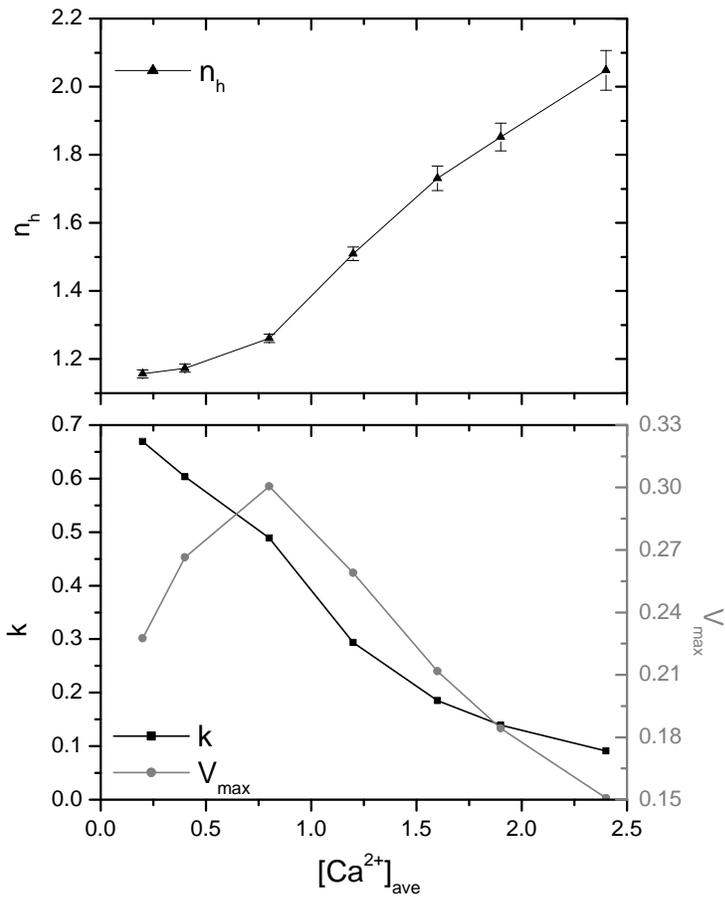}
\caption{Top panel: Hill coefficient $n_h$ for IP$_3$-induced
calcium release versus [Ca$^{2+}$]$_{ave}$, obtained fitting data
displayed in figure 2. Bottom panel: $k$ and $V_{max}$ values of
the Hill curves for IP$_3$-induced calcium release versus
[Ca$^{2+}$]$_{ave}$, obtained fitting data displayed in Fig. 2.}
\end{figure}

\begin{figure}[ht]
\includegraphics[width=10cm]{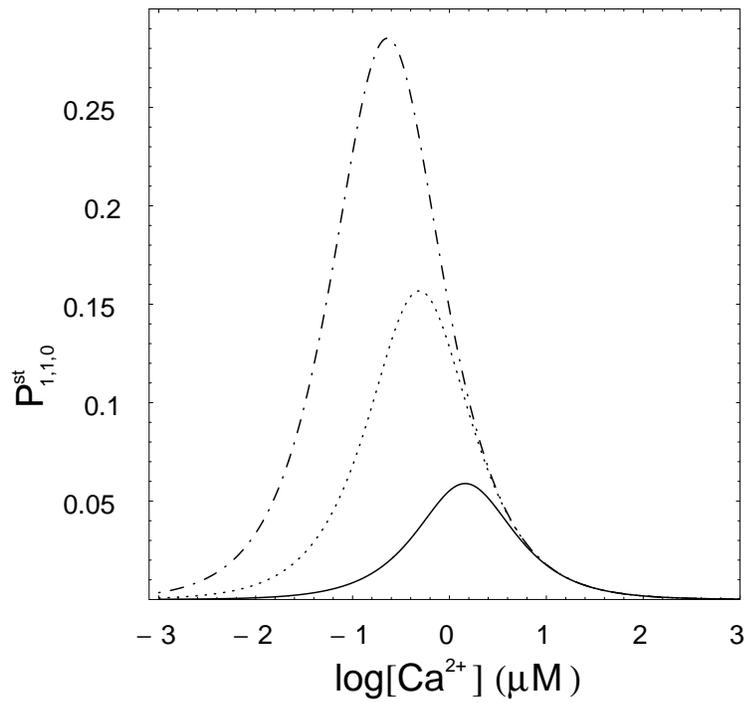}
\caption{The theoretically predicted probability to find the open channel
as a function of cytosolic [Ca$^{2+}$] for three different values of IP$_3$
concentration (0.01 $\mu$M solid line, 0.1 $\mu$M dotted line,
and 1$\mu$M dot-dashed line).}
\end{figure}

\begin{figure}[ht]
\includegraphics[width=15cm]{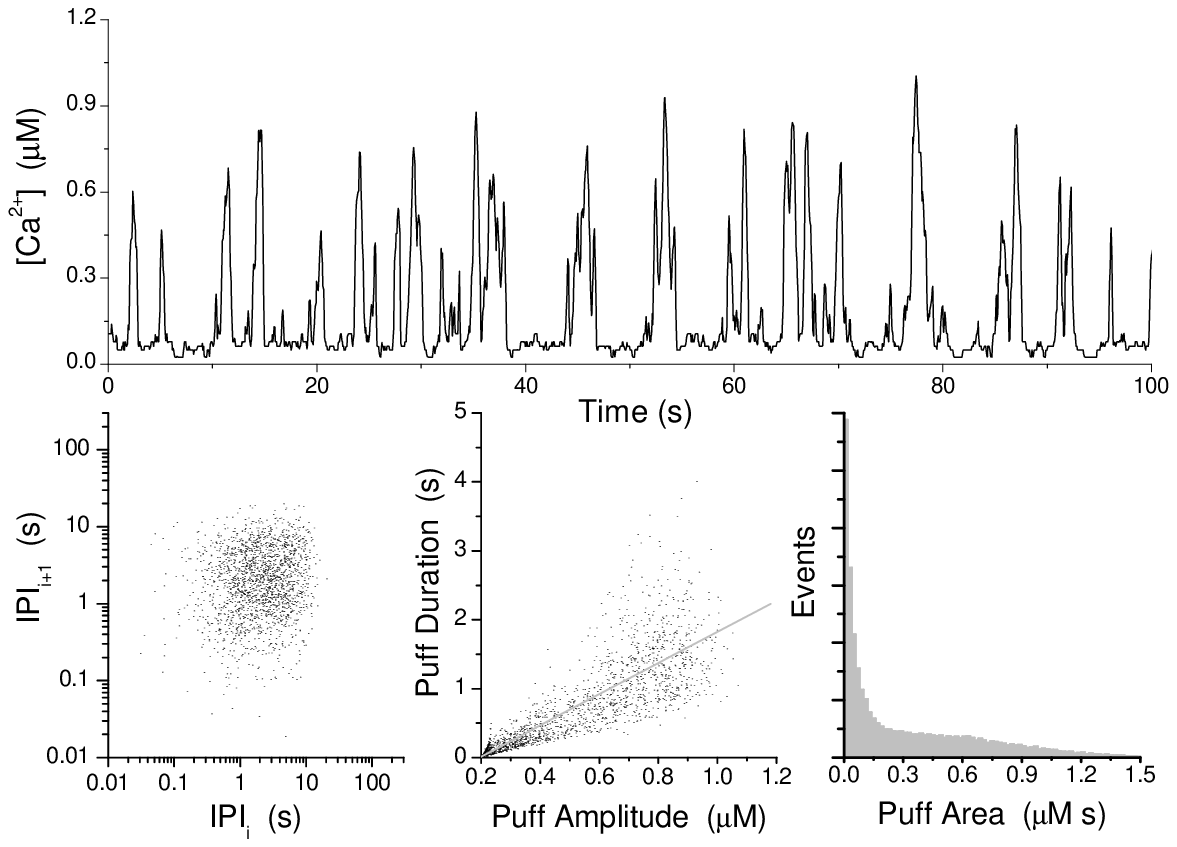}
\caption{Top panel: 100 seconds of the stochastic temporal
behavior of the calcium release from a cluster with 20 IP$_3$Rs
($N=20$, [Ca$^{2+}$]$_{ave}$=1.6 and IP$_3$=0.35). Bottom-left
panel: scatter plot of IPI$(i)$ versus IPI$(i+1)$ shown that there
is not temporal correlation between two consecutives puffs.
Bottom-center panel: scatter plot of puff amplitude versus puff
duration shown that there is some correlation between puff
durations and amplitudes of puffs. Bottom-right panel: puff area
distribution shown that there is Ca$2+$ release of all sizes. All
plots in the bottom panel were computed with puffs which
amplitudes exceeds the threshold 0.20 $\mu$M, recorded when
simulate 10,000 seconds.}
\end{figure}

\begin{figure}[ht]
\includegraphics[width=15cm]{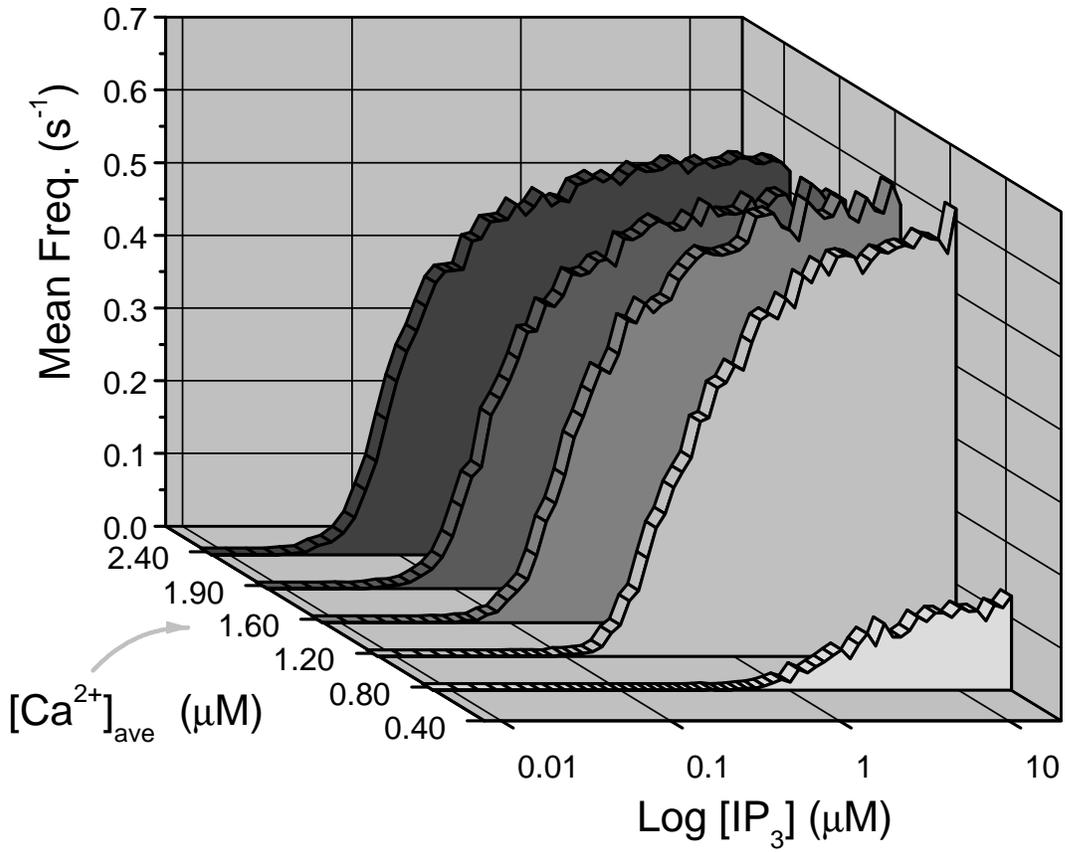}
\caption{Mean frequency versus IP$_3$ concentration, for five
different values of [Ca$^{2+}$]$_{ave}$. All plots were computed
with puffs which amplitudes exceeds the threshold 0.20 $\mu$M,
recorded when simulate 10,000 seconds.}
\end{figure}

\begin{figure}[ht]
\includegraphics[width=15cm]{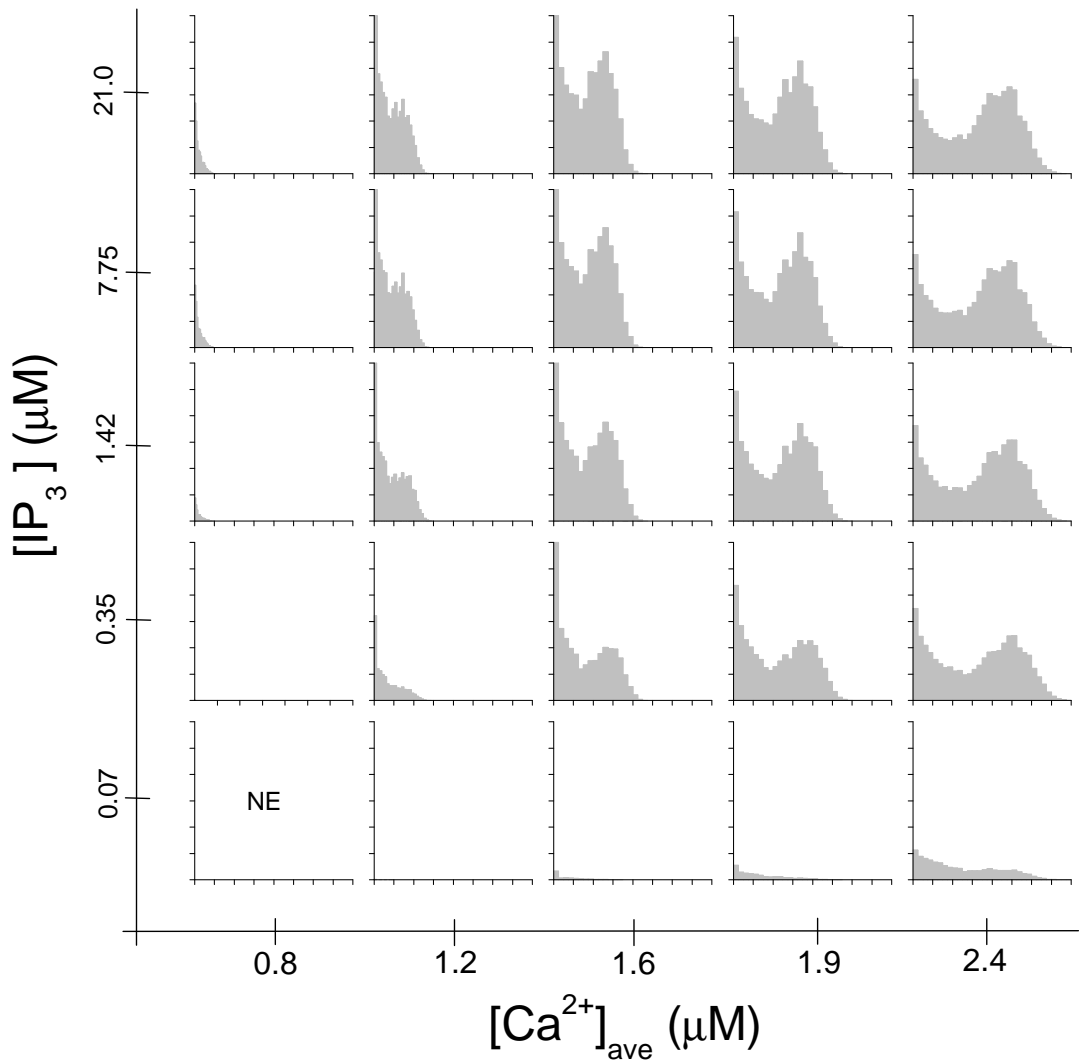}
\caption{Diagram of the puff amplitude distributions is shown in
the [Ca$^{2+}$]$_{ave}$-IP$_3$ plane. The distributions was
computed with all puffs, which amplitudes exceeds the threshold
0.20 $\mu$M, recorded in 10,000 seconds. In this condition no
events (NE) were recorded for [IP$_3$]= 0.07 $\mu$M and
[Ca$^{2+}$]$_{ave}$ = 0.8 $\mu$M. The range of the axis are the
same for all distributions, between [0.20,1.8] $\mu$M for the
amplitudes of puffs (horizontal axe), and between [0,6000] for the
number of events (vertical axe).}
\end{figure}

\begin{figure}[ht]
\includegraphics[width=15cm]{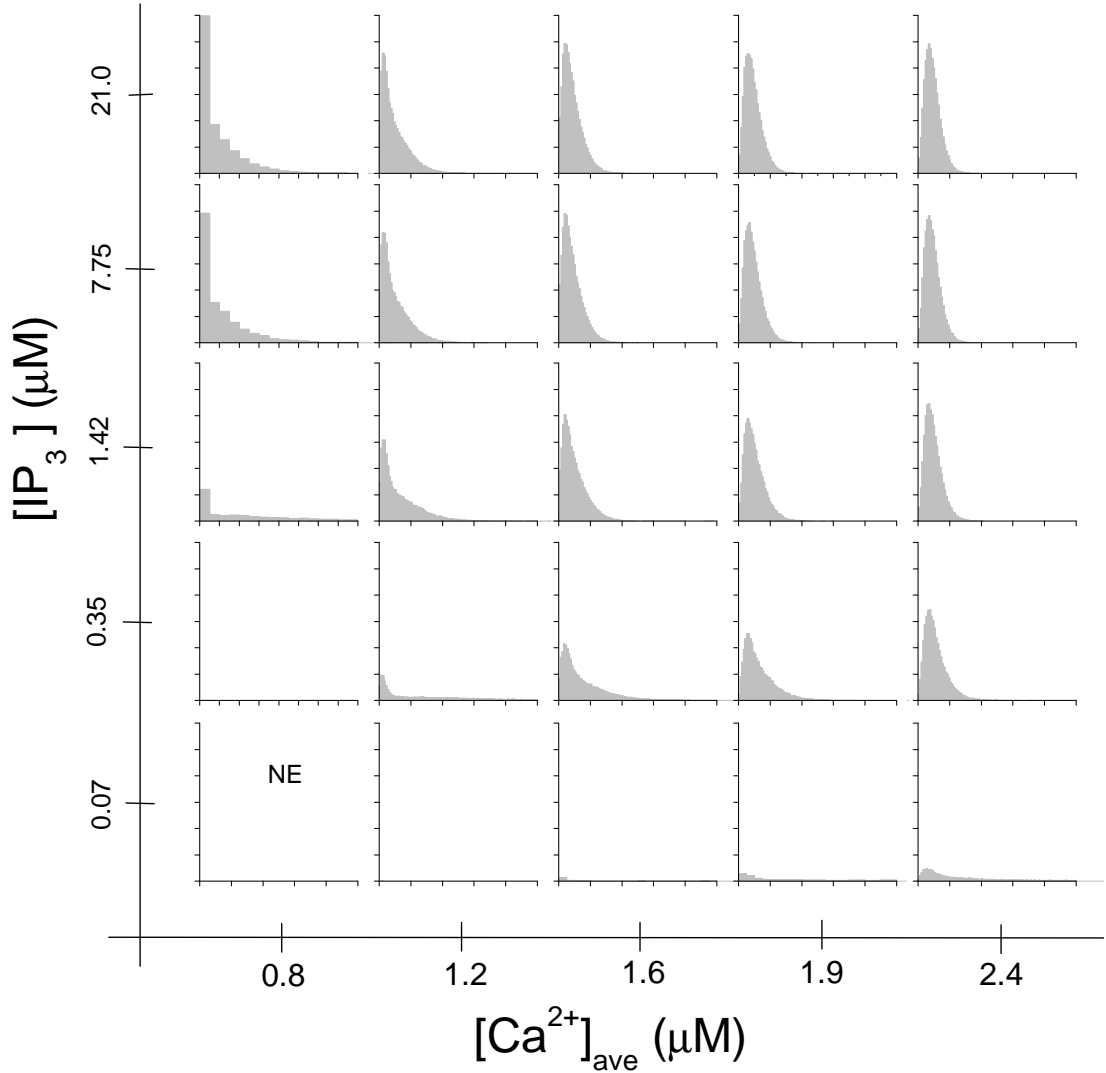}
\caption{Diagram of the inter puff interval (IPI) distributions is
shown in the [Ca$^{2+}$]$_{ave}$-IP$_3$ plane. The distributions
was computed with all puffs, which amplitudes exceeds the
threshold 0.20 $\mu$M, recorded in 10000 seconds. In this
condition no events (NE) were recorded for [IP$_3$]= 0.07 $\mu$M
and [Ca$^{2+}$]$_{ave}$ = 0.8 $\mu$M. The range of the vertical
axis are the same for all distributions, between [0,6000] events.
While the ranges of IPI (horizontal axis) are between [0,80]
seconds for the first column ([Ca$^{2+}$]$_{ave}$=0.8 $\mu$M), and
between [0,20] seconds in all other cases.}
\end{figure}

\begin{figure}[ht]
\includegraphics[width=15cm]{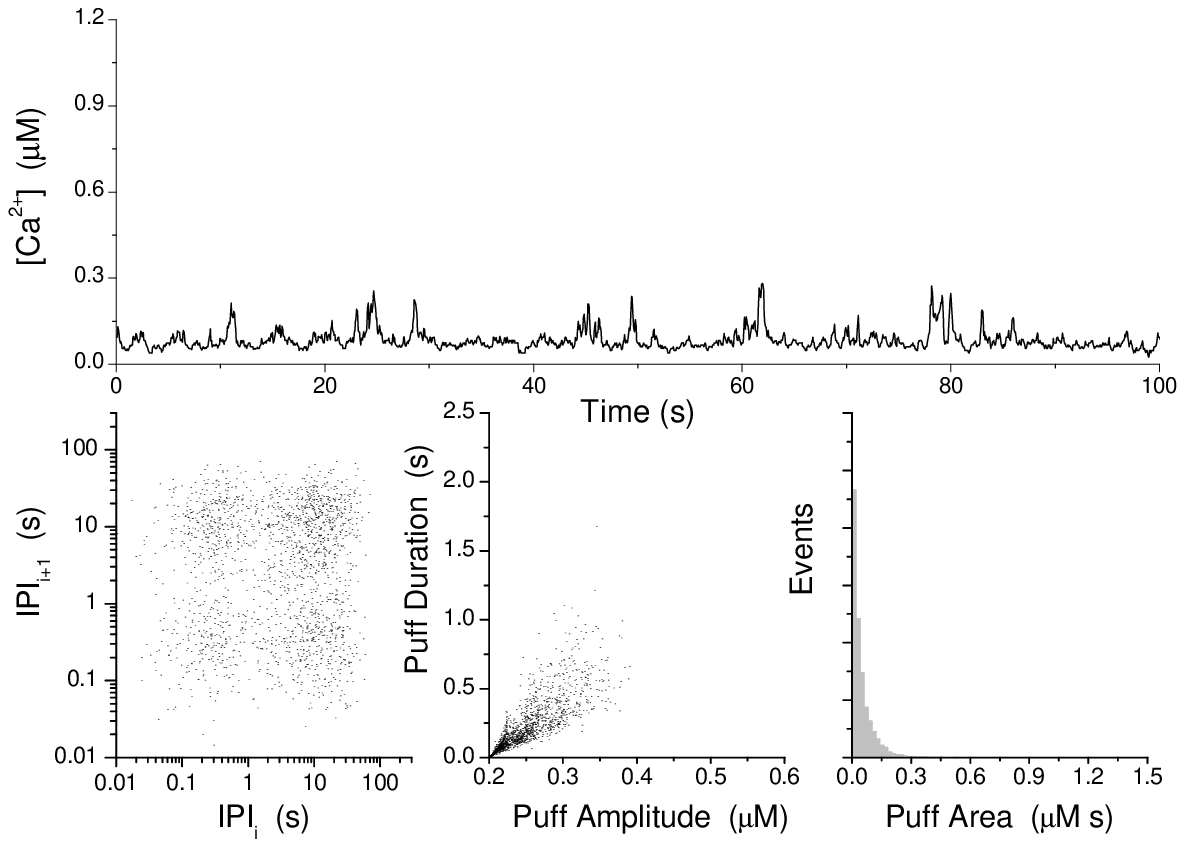}
\caption{Top panel: 100 seconds of the stochastic temporal
behavior of the calcium release from a cluster with 20 IP$_3$Rs
($N=20$, [Ca$^{2+}$]$_{ave}$=0.8 $\mu$M and IP$_3$=21.0 $\mu$M).
Bottom-left panel: scatter plot of IPI$(i)$ versus IPI$(i+1)$
shown that there is not temporal correlation between two
consecutives puffs. Bottom-center panel: scatter plot of puff
amplitude versus puff duration shown that there is some
correlation between puff durations and amplitudes of puffs.
Bottom-right panel: puff area distribution shown that there is
Ca$^{2+}$ release of all sizes. All plots in the bottom panel were
computed with puffs which amplitudes exceeds the threshold 0.20
$\mu$M, recorded in 10,000 seconds.}
\end{figure}

\begin{figure}[ht]
\includegraphics[width=15cm]{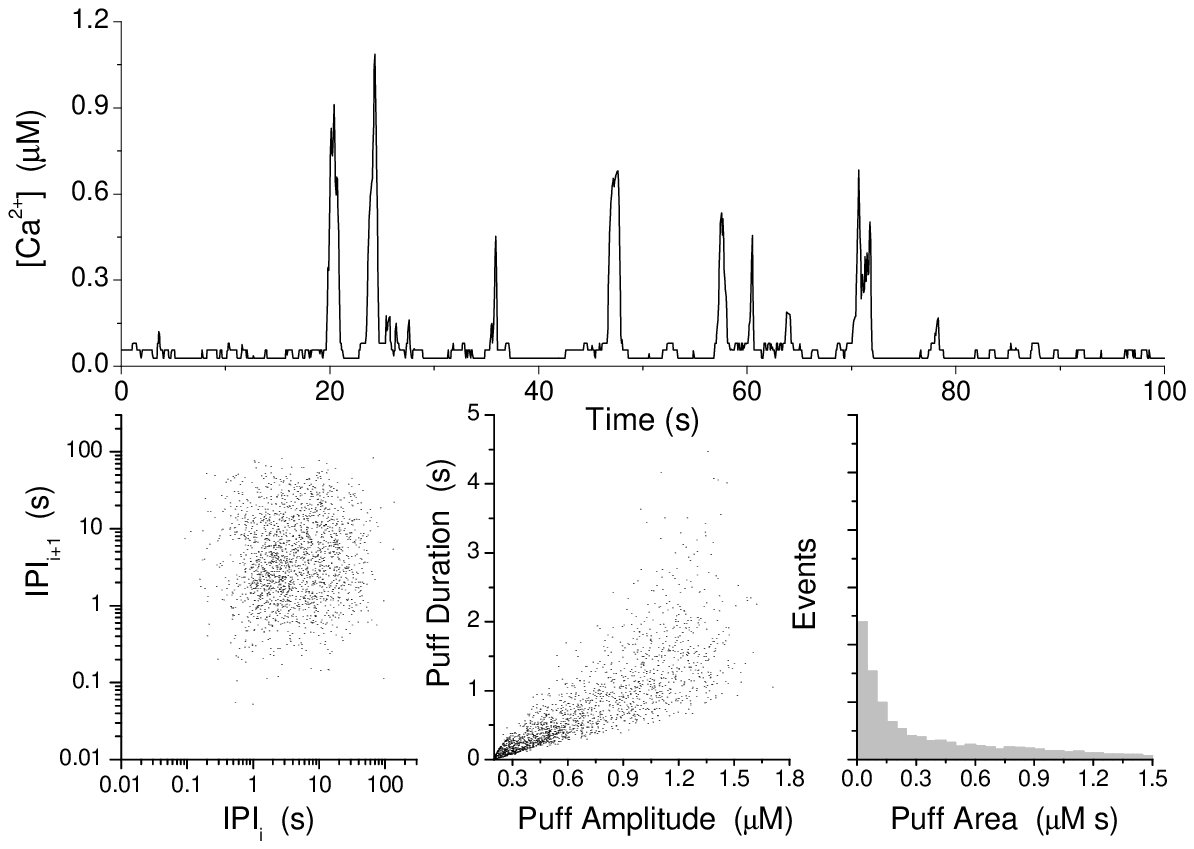}
\caption{Top panel: 100 seconds of the stochastic temporal
behavior of the calcium release from a cluster with 20 IP$_3$Rs
($N=20$, [Ca$^{2+}$]$_{ave}$=2.4 $\mu$M and IP$_3$=0.07 $\mu$M).
Bottom-left panel: scatter plot of IPI$(i)$ versus IPI$(i+1)$
shown that there is not temporal correlation between two
consecutives puffs. Bottom-center panel: scatter plot of puff
amplitude versus puff duration shown that there is some
correlation between puff durations and amplitudes of puffs.
Bottom-right panel: puff area distribution shown that there is
Ca$^{2+}$ release of all sizes. All plots in the bottom panel were
computed with puffs which amplitudes exceeds the threshold 0.20
$\mu$M, recorded in 10,000 seconds.}
\end{figure}

\end{document}